\documentclass[conference,10pt]{IEEEtran}
\IEEEoverridecommandlockouts
\usepackage{amsmath,amsfonts,amsthm,amssymb,bbm}
\usepackage{cite}
\usepackage{balance}
\usepackage{mathrsfs}
\usepackage{xcolor}

\DeclareMathOperator*{\argmax}{arg\,max}

\theoremstyle{plain} 
\newtheorem{theorem}{Theorem}
\newtheorem{corollary}{Corollary}
\newtheorem{definition}{Definition}

\newtheorem{proposition}{Proposition}
\theoremstyle{definition} \newtheorem{remark}{Remark}
\theoremstyle{definition}

\title{A Variational Formula for Infinity-R\'{e}nyi Divergence with Applications to Information Leakage}
\author{Gowtham R. Kurri, Oliver Kosut, Lalitha Sankar
\thanks{The authors are with the School of Electrical, Computer and Energy Engineering at Arizona State University. Email: {\tt gkurri@asu.edu, okosut@asu.edu, lalithasankar@asu.edu}}
\thanks{This work is supported in part by NSF grants CIF-1901243, CIF-1815361, CIF-2007688, CIF-2134256, and CIF-2031799.}
}
\begin{document}
\maketitle
\begin{abstract}
    We present a variational characterization for the R\'{e}nyi divergence of order infinity. Our characterization is related to guessing: the objective functional is a ratio of maximal expected values of a gain function applied to the probability of correctly guessing an unknown random variable. An important aspect of our variational characterization is that it remains agnostic to the particular gain function considered, as long as it satisfies some regularity conditions. Also, we define two variants of a tunable measure of information leakage, the maximal $\alpha$-leakage, and obtain closed-form expressions for these information measures by leveraging our variational characterization. 
\end{abstract}

\section{Introduction}\label{ection:intro}
R\'{e}nyi divergence was introduced by R\'{e}nyi~\cite{renyi1961measures} to quantify a measure of \emph{distance} between two probability distributions. It is parameterized by $\alpha$, called its order. It is closely tied to R\'{e}nyi entropy~\cite{renyi1961measures} in the same way as the Kullback-Leibler divergence (which is the R\'{e}nyi divergence of order $1$) is tied to the Shannon entropy. The R\'{e}nyi divergence has numerous applications in information theory and related fields; this includes hypothesis testing~\cite{morales2000renyi}, the multiple source adaptation problem~\cite{MansourMR09}, cryptography~\cite{bai2018improved}, uncertainty analysis of rare events~\cite{dupuis2020sensitivity} (see \cite{VanH14} for more applications). 

A variational characterization for a divergence transforms its definition into an optimization problem. Variational characterizations for R\'{e}nyi divergences of order $\alpha\in\mathbb{R}\setminus \{0\}$ are studied in the literature~\cite{Shayevitz11,VanH14,Sason16,Anantharam,BirrellDKRW21,pantazis2020cumulant}. In addition to being compelling mathematical tools to analyze probabilistic models, these characterizations have applications in hypothesis testing, divergence estimation from the data, generative adversarial networks (GANs), etc. In particular, Shayevitz~\cite{Shayevitz11}, van Erven and Harremo\"{e}s~\cite{VanH14}, and Anantharam~\cite{Anantharam} study a variational characterization where the objective functional is a linear combination of relative entropies, thereby offering a new operational interpretation for R\'{e}nyi divergence in a two-sensor composite hypothesis testing framework~\cite{Shayevitz11}. Birrell \emph{et al.}~\cite{BirrellDKRW21} study a variational characterization where the objective functional involves exponential integrals of bounded measurable functions and efficiently estimate the R\'{e}nyi divergence from the data. This characterization was later used to formulate the two-player min-max game in Cumulant GAN~\cite{pantazis2020cumulant}.

    The focus of this paper is on variational characterization for R\'{e}nyi divergence of order $\infty$, $D_\infty(P_X\|Q_X)=\log\max_{x\in\mathcal{X}}\frac{P_X(x)}{Q_X(x)}$. This divergence naturally shows up in the literature on \emph{minimum description length principle} in statistics as the \emph{worst-case regret} of mismatched coding~\cite{grunwald2007minimum} and is also related to the \emph{separation distance} used to bound the rate of convergence to the stationary distribution for certain Markov chains~\cite{aldous1987strong}. More recently, R\'{e}nyi quantities of order~$\infty$ are also explored in the context of the entropy-power inequality (EPI)~\cite{XuMM17} and common information~\cite{YuT18}. Even though the variational characterizations for R\'{e}nyi divergence mentioned earlier are presented for any finite order $\alpha$, one can obtain such characterizations for $\infty$-R\'{e}nyi divergence by applying a limiting argument (see the discussion above Proposition~\ref{proposition}). We develop a new variational characterization for $\infty$-R\'{enyi} divergence, specifically, we prove
\begin{align}\label{eqn:intro-varia}
    D_\infty(P_X||Q_X)=\sup_{P_{U|X}}\log{\frac{\sup_{P_{\hat{U}}}\mathbb{E}_{U\sim P_U}\left[g(P_{\hat{U}}(U))\right]}{\sup_{P_{\hat{U}}}\mathbb{E}_{U\sim Q_U}\left[g(P_{\hat{U}}(U))\right]}},
\end{align}
where $P_U(u)=\sum_xP_X(x)P_{U|X}(u|x)$, $Q_U(u)=\sum_xQ_X(x)P_{U|X}(u|x)$, and  $g:[0,1]\rightarrow \mathbb{R}_+$ is an arbitrary \emph{gain} function satisfying some mild assumptions; see Theorem~\ref{theorem:main-varchar}. The expressions in the ratio in \eqref{eqn:intro-varia} capture the maximal expected gains in guessing an unknown random variable (RV) $U$ distributed according to $P_U$ or $Q_U$, respectively. In a way, this ratio compares the distributions $P_X$ and $Q_X$ and is certainly dependent on the gain function $g$. However, our variational characterization in \eqref{eqn:intro-varia} shows that this ratio when optimized over all the channels $P_{U|X}$ remains constant irrespective of the gain function used. Our characterization differs from earlier characterizations in view of its connection to guessing and, more importantly, because of its robustness to the gain function. 


We also explore the connection and application of our variational characterization to information leakage measures. Recently, Issa~\emph{et al.}~\cite{Issaetal} introduced the measures, \emph{maximal leakage} (MaxL) and \emph{maximal realizable leakage} (MaxRL). For a given distribution $P_{XY}$, noting that $D_\infty(P_{XY}\|P_X\times P_Y)$ equals the MaxRL~\cite[Theorem~13]{Issaetal}, our variational characterization for this divergence complements that of the MaxL in terms of gain functions~\cite[Theorem~5]{Issaetal},\cite[Theorem~10]{AlvimCMMPS14}.  MaxL was later generalized in~\cite{LiaoKS20} to a family of leakages, \emph{maximal $\alpha$-leakage} (Max-$\alpha$L), that allows tuning the measure to specific applications. We now define two variants of Max-$\alpha$L, namely opportunistic maximal- and maximal realizable-$\alpha$ leakage, and obtain closed-form expressions for them (Theorems~\ref{thm:oppor-alpha-leakage} and \ref{thm:realiz-alpha-leakage}) using 
our variational characterization.

\section{A Variational Characterization of R\'{e}nyi Divergence of Order Infinity}
We begin by reviewing the definition of R\'{e}nyi divergence.
\begin{definition}(R\'{e}nyi divergence of order $\alpha$~\cite{renyi1961measures})
The R\'{e}nyi divergence of order $\alpha\in(0,1)\cup(1,\infty)$ between two probability distributions $P_X$ and $Q_X$ on a finite alphabet $\mathcal{X}$ is defined as
\begin{align}\label{eqn:Renyidiv-def}
    D_\alpha(P_X||Q_X)=\frac{1}{\alpha-1}\log\left(\sum_{x\in\mathcal{X}}P_X(x)^\alpha Q_X(x)^{1-\alpha}\right).
\end{align}
It is defined by its continuous extension for $\alpha=1$ and $\alpha=\infty$, respectively, and is given by
\begin{align}
    D_1(P_X||Q_X)&=\sum_{x\in\mathcal{X}}P_X(x)\log\frac{P_X(x)}{Q_X(x)},\\
    D_\infty(P_X||Q_X)&=\max_{x\in\mathcal{X}}\log\frac{P_X(x)}{Q_X(x)}.
\end{align}
\end{definition}
We present our main result below.
\begin{theorem}[A variational characterization for $D_\infty(\cdot||\cdot)$]\label{theorem:main-varchar}
Given two probability distributions $P_X$ and $Q_X$ on a finite alphabet $\mathcal{X}$, let $g:[0,1]\rightarrow [0,\infty)$ be a function satisfying the following assumptions:
\begin{itemize}
    \item $g(0)=0$ and $g$ is continuous at 0,
    \item $0<\sup_{p\in[0,1]}g(p)<\infty$.
\end{itemize}
Then, we have
\begin{align}\label{eqn:variationalmain}
    D_\infty(P_X||Q_X)=\sup_{P_{U|X}}\log{\frac{\sup_{P_{\hat{U}}}\mathbb{E}_{U\sim P_U}\left[g(P_{\hat{U}}(U))\right]}{\sup_{P_{\hat{U}}}\mathbb{E}_{U\sim Q_U}\left[g(P_{\hat{U}}(U))\right]}},
\end{align}
where $P_U(u)=\sum_xP_X(x)P_{U|X}(u|x)$ and $Q_U(u)=\sum_xQ_X(x)P_{U|X}(u|x)$.
\end{theorem}
\begin{remark}
Interestingly, there are non-positive gain functions too that do not satisfy the conditions in Theorem~\ref{theorem:main-varchar} but \eqref{eqn:variationalmain} still holds. For example, $g(t)=\log{t}$ is one such function (see Appendix~\ref{appendix-loggain} for details). 
\end{remark}
The proof of Theorem~\ref{theorem:main-varchar} is in Section~\ref{proof:mainresult}. Some examples of gain function $g$ that satisfy the conditions in Theorem~\ref{theorem:main-varchar} are
\begin{align*}
    g(p)=p^2,\ \mathbbm{1}\{p=1/2\}, \ \frac{\alpha}{\alpha-1}p^{\frac{\alpha-1}{\alpha}}, \ \text{where}\ \alpha\in(1,\infty).
\end{align*}
We obtain the following corollary from Theorem~\ref{theorem:main-varchar} by substituting the latter gain function 
$g_\alpha(t)=\frac{\alpha}{\alpha-1}t^{\frac{\alpha-1}{\alpha}}, \ \text{where}\ \alpha\in(1,\infty)$
(related to a class of adversarial loss functions, namely, $\alpha$-loss~\cite{LiaoKS20}) and using \cite[Lemma~1]{LiaoKS20} which gives closed-form expressions for the corresponding optimization problems in the numerator and the denominator in \eqref{eqn:variationalmain}.
\begin{corollary}\label{corollary:var3}
Given two probability distributions $P_X$ and $Q_X$ on a finite alphabet $\mathcal{X}$, we have, for $\alpha\in(1,\infty)$,
\begin{align}\label{prop1eq}
    D_\infty(P_X||Q_X)=\sup_{P_{U|X}}\log\frac{\left(\sum_uP_U(u)^\alpha\right)^\frac{1}{\alpha}}{\left(\sum_uQ_U(u)^\alpha\right)^\frac{1}{\alpha}},
\end{align}
where $P_U(u)=\sum_xP_X(x)P_{U|X}(u|x)$ and $Q_U(u)=\sum_xQ_X(x)P_{U|X}(u|x)$.
\end{corollary}
As mentioned earlier, we note that the existing variational characterizations for $D_\alpha(\cdot\|\cdot)$ (with finite $\alpha$)  also give rise to variational characterizations for $D_\infty(\cdot\|\cdot)$ by taking limit $\alpha\rightarrow \infty$. Shayevitz~\cite{Shayevitz11} and Birrell~\emph{et al.}~\cite{BirrellDKRW21} proved that 
\begin{align}
    D_\alpha(P_X\|Q_X)=\sup_{R_X:R_X\ll P_X}(D(R_X\|Q_X)-\frac{\alpha D(R_X\|P_X)}{\alpha-1})\label{eqn:varshayvitz}
    \end{align}
    and
\begin{align}
   &D_\alpha(P_X\|Q_X)\nonumber\\
   & =\sup_{g:\mathcal{X}\rightarrow \mathbb{R}}\left(\frac{\alpha\log{\mathbb{E}_{X\sim Q_X}\mathrm{e}^{(\alpha-1)g(X)}}}{\alpha-1}-\log{\mathbb{E}_{X\sim P_X}\mathrm{e}^{\alpha g(X)}}\right)\label{eqn:varBirrell},
\end{align}
respectively (more general forms of \eqref{eqn:varshayvitz} appear in \cite{VanH14,Sason16,Anantharam}). One can obtain the variational characterizations for $D_\infty(\cdot\|\cdot)$ by taking the limit $\alpha\rightarrow \infty$ in \eqref{eqn:varshayvitz} and assuming interchangeability of the limit and the supremum; 
one can similarly do so, in \eqref{eqn:varBirrell}, using a change of variable $f=\mathrm{e}^{\alpha g}$ and assuming interchangeability of the limit and the supremum. For the sake of completeness and rigor, we summarize the resulting variational forms for $D_\infty(\cdot\|\cdot)$ in the following proposition and present a proof in Appendix~\ref{appendix1}.
\begin{proposition}\label{proposition}
Given two probability distributions $P_X$ and $Q_X$ on a finite alphabet $\mathcal{X}$, we have
\begin{align}
     D_{\infty}(P_X\|Q_X)&=\sup_{R_X:R_X\ll P_X} \left(D(R_X\|Q_X)-D(R_X\|P_X)\right),\label{eqn:varinf-Anantharam}\\
      D_\infty(P_X\|Q_X)&= \sup_{f:\mathcal{X}\rightarrow[0,\infty)}\log\frac{\mathbb{E}_{X\sim P_X}[f(X)]}{\mathbb{E}_{X\sim Q_X}[f(X)]}\label{eqn:varinf-Birrell}.
\end{align}
\end{proposition}
\section{Applications to Information Leakage Measures}\label{section:leakagemeasures}
The leakage measures maximal leakage~\cite{Issaetal} and maximal $\alpha$-leakage~\cite{LiaoKS20} (including its variants defined here) can be expressed in terms of the Sibson mutual information.
\begin{definition}[Sibson mutual information of order $\alpha$~\cite{sibson1969information}]
For a given joint distribution $P_{XY}$ on finite alphabet $\mathcal{X}\times\mathcal{Y}$, the Sibson mutual information of order $\alpha\in(0,1)\cup(1,\infty)$ is 
\begin{align*}\label{eqn:sibson-def}
    I_\alpha^{\emph{S}}(S;Y)=\frac{\alpha}{\alpha-1}\log\sum_{y\in\mathcal{Y}}\left(\sum_{x\in\mathcal{X}}P_X(x)P_{Y|X}(y|X)^\alpha\right)^\frac{1}{\alpha}.
\end{align*}
It is defined by its continuous extension for $\alpha=1$ and $\alpha=\infty$, respectively, and is given by
\begin{align}
    I_1^\emph{S}(X;Y)&=I(X;Y) \ \emph{(Shannon mutual information)},\\
    I_\infty^\emph{S}(X;Y)&=\log\sum_{y\in\mathcal{Y}}\max_{x:P_X(x)>0}P_{Y|X}(y|x).
\end{align}
\end{definition}

 
We now review maximal $\alpha$-leakage~\cite{LiaoKS20}, and define some variants of it.
 \begin{definition}[Maximal $\alpha$-leakage~\cite{LiaoKS20}]
 Given a joint distribution $P_{XY}$ on finite alphabet $\mathcal{X}\times\mathcal{Y}$, for $\alpha\in(1,\infty)$, the maximal $\alpha$-leakage from $X$ to $Y$ is defined as $\mathcal{L}_\alpha^{\emph{max}}(X\rightarrow Y)=$
 \begin{align}
    \sup_{U:U-X-Y}\frac{\alpha}{\alpha-1}\log\frac{\sum_{y\in\emph{supp}(Y)}P_Y(y)\left(\sum_uP_{U|Y}(u|y)^\alpha\right)^{\frac{1}{\alpha}}}{\left(\sum_uP_{U}(u)^\alpha\right)^{\frac{1}{\alpha}}}.
 \end{align}

 \end{definition}
 Maximal $\alpha$-leakage captures the information leaked about \emph{any function} of the random variable $X$ to an adversary that observes a correlated random variable $Y$. Liao~\emph{et al.}~\cite{LiaoKS20} showed that 
 \begin{align}\label{eqn:maximalalpha}
     \mathcal{L}_\alpha^{\text{max}}(X\rightarrow Y)=\sup_{P_{\tilde{X}}}I_\alpha^{\text{S}}(\tilde{X};Y),
 \end{align}
 where the supremum is over all the probability distributions $P_{\tilde{X}}$ on the support of $P_X$. Maximal $\alpha$-leakage recovers maximal leakage~\cite{Issaetal}, another measure of information leakage, when $\alpha\rightarrow\infty$. 
 
 Motivated by Issa~\emph{et al.}~\cite[Definitions~2 and 8]{Issaetal}, we define the following variants of maximal $\alpha$-leakage depending on the type of the adversary. In particular, the definition of maximal $\alpha$-leakage corresponds to an adversary interested in a single randomized function of $X$. However, in some scenarios, the adversary could choose the guessing function depending on the realization of $Y$, leading to the following definition.
 \begin{definition}[Opportunistic maximal $\alpha$-leakage]
 Given a joint distribution $P_{XY}$ on a finite alphabet $\mathcal{X}\times\mathcal{Y}$, for $\alpha\in(1,\infty)$, the opportunistic maximal $\alpha$-leakage is $\tilde{\mathcal{L}}_\alpha^{\emph{max}}(X\rightarrow Y)=$
 \begin{align}\label{eqn:opportunistic}
 \frac{\alpha}{\alpha-1}\log \sum_{y\in\emph{supp}(Y)}P_Y(y)\sup_{U:U-X-Y}\frac{\left(\sum_uP_{U|Y}(u|y)^\alpha\right)^\frac{1}{\alpha}}{\left(\sum_uP_U(u)^\alpha\right)^{\frac{1}{\alpha}}}.
 \end{align}
 \end{definition}
  
 Maximal $\alpha$-leakage captures the \emph{average} (over $\mathcal{Y}$) guessing performance of the adversary. In some scenarios, it might be relevant to consider the \emph{maximum} instead of the average.
 \begin{definition}[Maximal realizable $\alpha$-leakage]
 Given a joint distribution $P_{XY}$ on a finite alphabet $\mathcal{X}\times\mathcal{Y}$, for $\alpha\in(1,\infty)$, the maximal realizable $\alpha$-leakage is $\mathcal{L}_\alpha^{\emph{r}-\max}(X\rightarrow Y)=$
 \begin{align}\label{eqn:realizable}
    \sup_{U:U-X-Y}\frac{\alpha}{\alpha-1}\log\frac{\max_{y\in\emph{supp}(Y)}\left(\sum_uP_{U|Y}(u|y)^\alpha\right)^{\frac{1}{\alpha}}}{\left(\sum_uP_U(u)^\alpha\right)^{\frac{1}{\alpha}}}.
 \end{align}

 \end{definition}

 Unlike the expression for maximal $\alpha$-leakage in \eqref{eqn:maximalalpha}, interestingly, it turns out that the closed-form expressions for the opportunistic maximal $\alpha$-leakage and maximal realizable $\alpha$-leakage do not explicitly depend on $\alpha$ (except via the scaling factor $\frac{\alpha}{\alpha-1}$). This is a consequence of the robustness of our variational characterization to the gain function (Corollary~\ref{corollary:var3}). We now present the closed-form expressions for these leakages.  

\begin{theorem}[Opportunistic maximal $\alpha$-leakage]\label{thm:oppor-alpha-leakage}
Given a joint distribution $P_{XY}$ on finite alphabet $\mathcal{X}\times\mathcal{Y}$, the opportunistic maximal $\alpha$-leakage, for $\alpha\in(1,\infty)$, is given by
\begin{align}\label{eqn:oppexp}
    \tilde{\mathcal{L}}_\alpha^{\emph{max}}(X\rightarrow Y)=\frac{\alpha}{\alpha-1}I_\infty^{\emph{S}}(X;Y).
\end{align}
\end{theorem}

\begin{theorem}[Realizable maximal $\alpha$-leakage]\label{thm:realiz-alpha-leakage}
Given a joint distribution $P_{XY}$ on finite alphabet $\mathcal{X}\times\mathcal{Y}$, the realizable maximal $\alpha$-leakage, for $\alpha\in(1,\infty)$, is given by
\begin{align}\label{eqn:realexp}
    \tilde{\mathcal{L}}_\alpha^{\emph{r}-\emph{max}}(X\rightarrow Y)
    =\frac{\alpha}{\alpha-1}D_{\infty}(P_{XY}\|P_X\times P_Y).
\end{align}
\end{theorem}
The proofs of Theorems~\ref{thm:oppor-alpha-leakage} and \ref{thm:realiz-alpha-leakage} are given in Sections~\ref{proof:thm2} and \ref{proof:thm3}, respectively. Theorems~\ref{thm:oppor-alpha-leakage} and \ref{thm:realiz-alpha-leakage} recover the expressions for opportunistic maximal leakage and maximal realizable leakage~\cite[Theorems~2 and 13]{Issaetal}, respectively, as $\alpha\rightarrow\infty$. 
Finally, it can be inferred from the above expressions that opportunistic maximal $1$-leakage and  maximal realizable $1$-leakage are both equal to $\infty$ as $\alpha\rightarrow 1$\footnote{Note that maximal $1$-leakage is equal to Shannon channel capacity and Shannon mutual information when we define it using the supremum first and the limit next, and the limit first and the supremum next, respectively~\cite[Theorem~2]{LiaoKS20}. We can show that the latter way of defining the opportunistic maximal $1$-leakage and the maximal realizable $1$-leakage also yields $\infty$.}.
\section{Proofs}
\subsection{Proof of Theorem~\ref{theorem:main-varchar}}\label{proof:mainresult}
We first prove the lower bound $\text{LHS}\geq\text{RHS}$. Consider
\begin{align}
&\sup_{P_{U|X}} \log \frac{\sup_{P_{\hat{U}}} \mathbb{E}_{U\sim P_U} g(P_{\hat{U}}(U))}{\sup_{P_{\hat{U}}} \mathbb{E}_{U\sim Q_U} g(P_{\hat{U}}(U))}\nonumber\\
&=\sup_{P_{U|X}} \log \frac{\sup_{P_{\hat{U}}} \sum_{x,u} P_X(x)P_{U|X}(u|x) g(P_{\hat{U}}(u))}{\sup_{P_{\hat{U}}} \sum_{x,u} Q_X(x)P_{U|X}(u|x) g(P_{\hat{U}}(u))}\\
&=\sup_{P_{U|X}} \sup_{P_{\hat{U}}} \inf_{Q_{\hat{U}}} \log \frac{\sum_{x,u} P_X(x)P_{U|X}(u|x) g(P_{\hat{U}}(u))}{\sum_{x,u} Q_X(x)P_{U|X}(u|x) g(Q_{\hat{U}}(u))}\\
&\le \sup_{P_{U|X}} \sup_{P_{\hat{U}}}\log \frac{\sum_{x,u} P_X(x)P_{U|X}(u|x) g(P_{\hat{U}}(u))}{\sum_{x,u} Q_X(x)P_{U|X}(u|x) g(P_{\hat{U}}(u))}\\
&\le \sup_{P_{U|X}} \sup_{P_{\hat{U}}}\max_{x:P_X(x)>0} \log \frac{P_X(x) \sum_{u} P_{U|X}(u|x) g(P_{\hat{U}}(u))}{Q_X(x)\sum_{u} P_{U|X}(u|x) g(P_{\hat{U}}(u))}\label{eqn:thm1proofratiofact}\\
&=\max_{x:P_X(x)>0} \log \frac{P_X(x)}{Q_X(x)}\\
&=D_\infty(P_X\|Q_X).
\end{align}
where \eqref{eqn:thm1proofratiofact} follows because $\frac{\sum_ia_i}{\sum_ib_i}\leq \max_i\frac{a_i}{b_i}$, for $b_i>0$, $\forall i$. 

Now we prove the upper bound $\text{LHS}\leq\text{RHS}$. We lower bound the RHS of \eqref{eqn:variationalmain} by choosing a specific ``shattered" $P_{U|X}$. We pick a letter $x^\star$, and let $\mathcal{U}=\{x^\star\}\uplus\biguplus_{x\ne x^\star}\mathcal{U}_{x}$, where $|\mathcal{U}_{x}|=m$ for each $x\ne x^\star$. Then define
\begin{align}
P_{U|X}(u|x)=\begin{cases} 1 & u=x=x^\star,\\ 1/m & u\in\mathcal{U}_x,x\ne x^\star, \\ 0 & \text{otherwise.}\end{cases}
\end{align}
Note that
\begin{align}
P_U(u)=\begin{cases} P_X(x^\star) & u=x^\star\\ P_X(x)/m, & u\in\mathcal{U}_x,x\ne x^\star\end{cases}, 
\end{align}
and
\begin{align}Q_U(u)=\begin{cases} Q_X(x^\star) & u=x^\star\\ Q_X(x)/m, & u\in\mathcal{U}_x,x\ne x^\star\end{cases}.
\end{align}
Consider the numerator of the objective function in the RHS of \eqref{eqn:variationalmain}. We have
\begin{align}
\sup_{P_{\hat{U}}} \mathbb{E}_{U\sim P_U} \left[g(P_{\hat{U}}(U))\right]
&=\sup_{P_{\hat{U}}} \sum_u P_U(u) g(P_{\hat{U}}(u))
\\&\geq \sup_{P_{\hat{U}}} P_X(x^\star) g(P_{\hat{U}}(x^\star))
\\&=P_X(x^\star) \sup_{q\in[0,1]} g(q)\label{eqn:thm1proof1}.
\end{align}
Note that the expression in \eqref{eqn:thm1proof1} is finite because of the assumption on $g$ that $\sup_{q\in[0,1]}g(q)<\infty$. 

To bound the denominator of the objective function in the RHS of \eqref{eqn:variationalmain}, we will need the upper concave envelope of $g$, denoted $g^{**}$. Since $g$ is a function of a scalar, its upper concave envelope can be written as
\begin{align}
g^{**}(q)=\sup_{a,b,\lambda\in[0,1]: a\lambda+b(1-\lambda)=q} \lambda g(a)+(1-\lambda) g(b).
\end{align}
We claim that $g^{**}(0)=0$ and $g^{**}$ is continuous at 0. Fix some $\epsilon>0$. It suffices to show that there exists $\delta>0$ where $g^{**}(q)\leq \epsilon$ for all $q\in[0,\delta]$. By the assumption that $g(0)=0$ and $g$ is continuous at 0, there exists a $\delta$ small enough so that $g(q)\leq \epsilon/2$ for all $q\in[0,\sqrt{\delta}]$. Now, for any $q\in[0,\delta]$, consider any $a,b,\lambda$ where $a\lambda+b(1-\lambda)=q$. We assume without loss of generality that $a\leq q\leq b$. If $b\leq\sqrt{\delta}$, then we have $\lambda g(a)+(1-\lambda) g(b)\leq \epsilon/2$.
If $b>\sqrt{\delta}$, then we have
\begin{align}
q=a\lambda+b(1-\lambda)\geq b(1-\lambda)>\sqrt{\delta}(1-\lambda).
\end{align}
So we get $1-\lambda<\frac{q}{\sqrt{\delta}}\leq \frac{\delta}{\sqrt{\delta}}\leq\sqrt{\delta}$.
Thus
\begin{align}
\lambda g(a)+(1-\lambda) g(b)
&\leq \epsilon/2+\sqrt{\delta} \sup_{q\in[0,1]} g(q)\leq\epsilon\label{eqn:thm1proof5},
\end{align}
where \eqref{eqn:thm1proof5} holds for sufficiently small $\delta$, and again we have used the assumption that $\sup_{q\in[0,1]} g(q)<\infty$. This proves that $g^{**}(q)\leq\epsilon$ whenever $q\in[0,\delta]$. In particular, for sufficiently large $m$, 
\begin{align}\label{eqn:proofthm13}
\sup_{q\in[0,1/m]} g^{**}(q)\leq \epsilon.
\end{align}
Now the denominator in \eqref{eqn:variationalmain} can be upper bounded as
\begin{align}
&\sup_{P_{\hat{U}}} \mathbb{E}_{U\sim Q_U} \left[g(P_{\hat{U}}(U)))\right]\nonumber\\
&=\sup_{P_{\hat{U}}} \sum_u Q_U(u) g(P_{\hat{U}}(u))\\
&=\sup_{P_{\hat{U}}} (Q_X(x^\star) g(P_{\hat{U}}(x^\star))+\sum_{x\ne x^\star} Q_X(x) \sum_{u\in\mathcal{U}_x} \frac{1}{m} g(P_{\hat{U}}(u)))\\
&\leq \sup_{P_{\hat{U}}} Q_X(x^\star) g(P_{\hat{U}}(x^\star))+\sum_{x\neq x^\star} Q_X(x)  g^{**}(\sum_{u\in\mathcal{U}_x} \frac{1}{m} P_{\hat{U}}(u))\label{eqn:thm1proof2}\\
&\leq \sup_{q\in[0,1]} Q_X(x^\star) g(q)+\sum_{x\neq x^\star} Q_X(x) \sup_{q\in[0,1/m]} g^{**}(q)\\
&\leq \sup_{q\in[0,1]} Q_X(x^\star) g(q)+(1-Q_X(x^\star)) \epsilon,\label{eqn:thm1proof3}
\end{align}
where \eqref{eqn:thm1proof2} follows from the definition of the upper concave envelope and \eqref{eqn:thm1proof3} follows from \eqref{eqn:proofthm13} for sufficiently large $m$.

Putting together the bounds in \eqref{eqn:thm1proof1} and \eqref{eqn:thm1proof3}, we have
\begin{align}
   & \sup_{P_{U|X}} \log \frac{\sup_{P_{\hat{U}}} \mathbb{E}_{U\sim P_U} g(P_{\hat{U}}(U))}{\sup_{P_{\hat{U}}} \mathbb{E}_{U\sim Q_U} g(P_{\hat{U}}(U))}\nonumber\\
    &\geq \log\max_{x^\star}\ \sup_{\epsilon>0}\ \frac{\sup_{q\in[0,1]} P_X(x^\star) g(q)}{\sup_{q\in[0,1]} Q_X(x^\star) g(q)+(1-Q_X(x^\star)) \epsilon}\\
        &= \log\max_{x^\star} \frac{\sup_{q\in[0,1]} P_X(x^\star) g(q)}{\sup_{q\in[0,1]} Q_X(x^\star) g(q)}\\
        &=\log\max_{x^\star} \frac{P_X(x^\star)}{Q_X(x^\star)}\label{eqn:thm1proof4}\\
        &=D_{\infty}(P_X\|Q_X),
\end{align}
where \eqref{eqn:thm1proof4} uses the assumption that $\sup_{q\in[0,1]}g(q)<\infty$. Finally, we note that the assumption $\sup_{q\in[0,1]}g(q)>0$ is to ensure that the objective function in \eqref{eqn:variationalmain} is well-defined. In particular, for any $P_{U|X}$, fix a $u^\prime$ such that $P_U(u^\prime)>0$. Then we have
\begin{align}
    \sup_{P_{\hat{U}}} \mathbb{E}_{U\sim P_U} \left[g(P_{\hat{U}}(U))\right]&\geq \sup_{P_{\hat{U}}} g(P_{\hat{U}}(u^\prime))P_U(u^\prime)\\
    &= P_U(u^\prime)\sup_{q\in[0,1]}g(q)>0.
\end{align}
Similarly,  $\sup_{P_{\hat{U}}} \mathbb{E}_{U\sim Q_U} \left[g(P_{\hat{U}}(U))\right]>0$.
\subsection{Proof of Theorem~\ref{thm:oppor-alpha-leakage}}\label{proof:thm2}
From the expression in \eqref{eqn:opportunistic}, we have
\begin{align}
&\tilde{\mathcal{L}}_\alpha^{\text{max}}(X\rightarrow Y)\nonumber\\
     &=\frac{\alpha}{\alpha-1}\log \sum_{y\in\text{supp}(Y)}P_Y(y)\sup_{U:U-X-Y}\frac{\left(\sum_uP_{U|Y}(u|y)^\alpha\right)^\frac{1}{\alpha}}{\left(\sum_uP_U(u)^\alpha\right)^{\frac{1}{\alpha}}}\\
     &=\frac{\alpha}{\alpha-1}\log\sum_{y:P_Y(y)>0}P_Y(y)\max_{x:P_{X|Y}(x|y)>0}\frac{P_{X|Y}(x|y)}{P_{X}(x)}\label{eqn:opportunisticproof2}\\
     &=\frac{\alpha}{\alpha-1}\log\sum_{y:P_Y(y)>0}P_Y(y)\max_{x:P_{X|Y}(x|y)>0}\frac{P_{Y|X}(x|y)}{P_{Y}(y)}\label{eqn:opportunisticproof1}\\
     &=\frac{\alpha}{\alpha-1}\log\sum_{y:P_Y(y)>0}\max_{x:P_{X}(x)>0}P_{Y|X}(y|x)\\
     &=\frac{\alpha}{\alpha-1}I_{\infty}^{\text{S}}(X;Y),
\end{align}
where \eqref{eqn:opportunisticproof2} follows from Corollary~\ref{corollary:var3} and \eqref{eqn:opportunisticproof1} follows from the Bayes' Rule. 

\subsection{Proof of Theorem~\ref{thm:realiz-alpha-leakage}}\label{proof:thm3}
From the expression in \eqref{eqn:realizable}, we have
\begin{align}
&\mathcal{L}_\alpha^{\text{r}-\max}(X\rightarrow Y)\nonumber\\
    &\sup_{U:U-X-Y}\frac{\alpha}{\alpha-1}\log\frac{\max_{y\in\text{supp}(Y)}\left(\sum_uP_{U|Y}(u|y)^\alpha\right)^{\frac{1}{\alpha}}}{\left(\sum_uP_U(u)^\alpha\right)^{\frac{1}{\alpha}}}\nonumber\\
    &=\frac{\alpha}{\alpha-1}\log\max_{y\in\text{supp}(Y)}\sup_{U:U-X-Y}\frac{\left(\sum_uP_{U|Y}(u|y)^\alpha\right)^{\frac{1}{\alpha}}}{\left(\sum_uP_U(u)^\alpha\right)^{\frac{1}{\alpha}}}\\
    &=\frac{\alpha}{\alpha-1}\log\max_{y:P_{Y}(y)>0}\max_{x:P_{X|Y}(x|y)>0}\frac{P_{X|Y}(x)}{P_{X}(x)}\label{eqn:realizableproof1}\\
    &=\frac{\alpha}{\alpha-1}\log\max_{(x,y):P_{XY}(x,y)>0}\frac{P_{XY}(x,y)}{P_X(x)P_Y(y)}\\
    &=\frac{\alpha}{\alpha-1}D_{\infty}(P_{XY}\|P_X\times P_Y),
\end{align}
where \eqref{eqn:realizableproof1} follows from Corollary~\ref{corollary:var3}. 


\appendices
\balance
\section{Variational Characterization for $D_\infty(P_X\|Q_X)$ with Gain Function $g(t)=\log{t}$}\label{appendix-loggain}
Here we show that \eqref{eqn:variationalmain} holds for the non-positive gain function $g(t)=\log{t}$ that does not satisfy the conditions in Theorem~\ref{theorem:main-varchar}. The proof of the lower bound follows exactly along the same lines as that of Theorem~\ref{theorem:main-varchar} with the only difference that \eqref{eqn:thm1proofratiofact} holds for negative gain functions too noticing that $\frac{\sum_ia_i}{\sum_ib_i}\leq \max_i\frac{a_i}{b_i}$, for $b_i<0$, $\forall i$. 

For the upper bound, we first note that 
\begin{align}
    \sup_{P_{\hat{U}}}\mathbb{E}_{U\sim P_U}\left[\log{P_{\hat{U}}(U)}\right]&=-\inf_{P_{\hat{U}}} \left(H_P(U)+D(P_U\|P_{\hat{U}})\right)\nonumber\\
    &=-H_P(U).
\end{align}
We lower bound the RHS in \eqref{eqn:variationalmain} with gain function $g(t)=\log{t}$ by choosing a specific ``shattered" $P_{U|X}$. Let $\mathcal{U}=\uplus_{x\in\mathcal{X}}\mathcal{U}_x, \ |\mathcal{U}_x|=m_x$.
Define $P_{U|X}(u|x)=\frac{1}{m_x}$, $u\in\mathcal{U}_x$. So, we have
\begin{align}
       &\sup_{P_{U|X}}\frac{\sup_{P_{\hat{U}}}\mathbb{E}_{U\sim P_U}\left[\log{P_{\hat{U}}(U)}\right]}{\sup_{P_{\hat{U}}}\mathbb{E}_{U\sim Q_U}\left[\log{Q_{\hat{U}}(U)}\right]}
       \geq\frac{-H_P(U)}{-H_Q(U)}\\
       &=\frac{\sum_u\left(\sum_xP_X(x)P_{U|X}(u|x)\right)\log{\left(\sum_xP_X(x)P_{U|X}(u|x)\right)}}{\sum_u\left(\sum_xQ_X(x)P_{U|X}(u|x)\right)\log{\left(\sum_xQ_X(x)P_{U|X}(u|x)\right)}}\\
       &=\frac{\sum_xP_X(x)\log{\frac{P_X(x)}{m_x}}}{\sum_xQ_X(x)\log{\frac{Q_X(x)}{m_x}}}\\
       &=\frac{-H_P(X)/\log{m_{x^*}}-P_X(x^*)}{-H_Q(X)/\log{m_{x^*}}-Q_X(x^*)}\label{eqn:log-var-1}\\
       &=\frac{P_X(x^*)}{Q_X(x^*)}\label{eqn:log-var2}\\
       &=2^{D_\infty(P_X\|Q_X)},
\end{align}
where \eqref{eqn:log-var-1} follows by fixing an $x^*\in\argmax_x\frac{P_X(x)}{Q_X(x)}$ and choosing $m_x=1$, for $x\neq x^*$, and \eqref{eqn:log-var2} then follows by taking limit $m_{x^*}\rightarrow \infty$.
\section{Proof of Proposition~\ref{proposition}}\label{appendix1}
 For any $R_X\ll P_X$, consider
\begin{align}
   & D(R_X\|Q_X)-D(R_X\|P_X)\nonumber\\
    &=\sum_{x}R_X(x)\log{\frac{R_X(x)}{Q_X(x)}}-\sum_{x}R_X(x)\log{\frac{R_X(x)}{P_X(x)}}\\
    &=\sum_xR_X(x)\log{\frac{P_X(x)}{Q_X(x)}}\\
    &\leq \sum_x R_X(x)\left(\max_{x^\prime}\log\frac{P_X(x^\prime)}{Q_X(x^\prime)}\right)\label{eqn:appendix1}\\
    &=\max_{x^\prime}\log\frac{P_X(x^\prime)}{Q_X(x^\prime)}\\
    &=D_\infty(P_X\|Q_X).
\end{align}
Moreover, for $R_X$ such that $R_X(x^*)=1$ for a fixed $x^*\in\argmax \frac{P_X(x)}{Q_X(x)}$, \eqref{eqn:appendix1} is tight. This proves \eqref{eqn:varinf-Anantharam}.

To prove \eqref{eqn:varinf-Birrell}, for the upper bound, we give a choice of the function $f$ for which the objective function in the RHS of \eqref{eqn:varinf-Birrell} is equal to $D_\infty(P_X||Q_X)$. In particular, fix an $x^*\in\argmax\frac{P_X(x)}{Q_X(x)}$ and consider a function $\tilde{f}$ defined by
\begin{align}
    \tilde{f}(x)=\begin{cases}
    1, &\ \text{if}\ x=x^*,\\
    0, &\ \text{otherwise}
    \end{cases}.
\end{align}
Clearly, we have 
\begin{align}\log\frac{\mathbb{E}_{X\sim P_X}[\tilde{f}(X)]}{\mathbb{E}_{X\sim Q_X}[\tilde{f}(X)]}=\log\frac{P_X(x^*)}{Q_X(x^*)}=D_\infty(P_X||Q_X).
\end{align}
For the lower bound, consider
\begin{align}
    \log\frac{ \mathbb{E}_{X\sim P_X}[f(X)]}{ \mathbb{E}_{X\sim Q_X}[f(X)]}&=\log\frac{\sum_xP_X(x)f(x)}{\sum_xQ_X(x)f(x)}\\
    &\leq \log\max_x\frac{P_X(x)f(x)}{Q_X(x)f(x)}\label{eqn:appendixvar2proof1}\\
    &=\max_x\log\frac{P_X(x)}{Q_X(x)}=D_\infty(P_X\|Q_X),
\end{align}
where \eqref{eqn:appendixvar2proof1} follows from the fact that $\frac{\sum_ia_i}{\sum_ib_i}\leq \max_i\frac{a_i}{b_i}$, for $b_i>0$, $\forall i$. 
Taking supremum over all $f$, we get
\begin{align}
    \sup_{f:\mathcal{X}\rightarrow[0,\infty)}\log\frac{\mathbb{E}_{X\sim P_X}[f(X)]}{\mathbb{E}_{X\sim Q_X}[f(X)]}&\leq \log\max_{x\in\mathcal{X}}\frac{P_X(x)}{Q_X(x)}\\
    &=D_\infty(P_X||Q_X).
\end{align}
This proves \eqref{eqn:varinf-Birrell}.
\bibliographystyle{IEEEtran}
\bibliography{Bibliography}
\end{document}